\documentclass[12pt]{article}
\title{
\bf The Abelian Higgs Model\\
as an Ensemble of Vortex Loops}  
\author{Dmitri Antonov 
\thanks{E-mail address: 
{\tt antonov@vxitep.itep.ru}}{\,} 
\thanks {Address after October 1999: INFN-Sezione di Pisa, 
Universit\'a degli studi di Pisa, Dipartimento di Fisica, 
Via Buonarroti, 2 - Ed. B - 56127 Pisa, Italy.} 
\\
{\it Institute of Theoretical and Experimental Physics,}\\
{\it B. Cheremushkinskaya 25, RU-117 218 Moscow, Russia}} 
\date{}
\begin{document}
\maketitle
\vspace{1mm}
\centerline{\bf Dedicated to Prof. Yu.A. Simonov on the occasion of his  
65-th birthday}
\vspace{10mm}
\centerline{\bf Abstract}
\vspace{1mm}
In the London limit of the 
Ginzburg-Landau theory (Abelian Higgs model), vortex dipoles (small vortex 
loops) are treated as a grand canonical ensemble 
in the dilute gas approximation.
The summation over these objects 
with the most general rotation- and translation invariant measure 
of integration over their shapes leads to effective sine-Gordon 
theories of the dual fields. 
The representations of the partition functions of 
both grand 
canonical ensembles are derived in the form of the integrals over 
the vortex dipoles and the small vortex loops, respectively.   
By virtue of these representations, 
the bilocal correlator of the vortex dipoles (loops) 
is calculated in the low-energy limit. 

It is further demonstrated that once the vortex dipoles  
(loops) are considered as such an ensemble rather than individual 
ones, the London limit of the 
Ginzburg-Landau theory (Abelian Higgs model) with external 
monopoles is equivalent up to the leading order in the inverse 
UV cutoff 
to the compact QED in the 
corresponding dimension with the charge of Cooper pairs changed due to the 
Debye screening.

\newpage

\section{Introduction}

It is commonly argued that Abrikosov vortices~\cite{abrikosov} 
(Nielsen-Olesen 
strings~\cite{nielsen}) in the dual 
Ginzburg-Landau theory 
(Abelian Higgs model) 
reveal the properties similar to those 
of strings in 3D- and 4D QCD, respectively. This observation 
is based on the so-called 't Hooft-Mandelstam scenario of 
confinement~\cite{thooft}, according to which confinement in QCD 
can be macroscopically thought of as the dual Meissner effect. 
Once being put forward, such a 
correspondence then enabled one to develop various 
phenomenological models of confinement, {\it e.g.} the 
so-called dual QCD approach~\cite{baker}. However, an extremely 
strong support has been given to the 't Hooft-Mandelstam 
scenario by the so-called Abelian projection 
method~\cite{abpr} (see {\it e.g.} Refs.~\cite{maedan, digiacomo, max,  
reinhardt, kondo, wipf, plb, epjnew, su2} for recent developments 
and Ref.~\cite{proc} for a review). The main outcome of this method 
is that under certain quite pleasurable assumptions (like the 
so-called Abelian dominance hypothesis~\cite{abdom, max}), 
$SU(2)$-QCD can indeed 
be with a good accuracy viewed as the London limit of the 
dual Abelian Higgs model with external electrically charged particles.  
An independent support of this conclusion comes out from the evaluation 
of field correlators in the dual Abelian Higgs model and comparison 
of them with those in QCD, introduced within the so-called 
Stochastic Vacuum Model~\cite{svm} and measured in the lattice 
experiments in Ref.~\cite{adriano}. 
This calculation has been performed in Refs.~\cite{correl, 
correl1, correl2}, and as a result a 
very good agreement between the two sets of correlators has 
been established~\footnote{A similar calculation of field 
correlators in the Abelian-projected $SU(3)$-gluodynamics has been 
performed in Ref.~\cite{plb}.}. All that encourages one to work further 
on investigating the confining properties of the Abelian Higgs model, which 
is a good and simple example of a 
model exhibiting the property of confinement. This is the main 
motivation for the present research. 

In a variety of an existing literature on the Abelian Higgs 
model from the point of view of the Abelian projection method 
(see {\it e.g.} Refs.~\cite{orl, axion, correl2}), the interaction of the 
Nielsen-Olesen strings has not been studied. In this respect, string 
representations of various objects (like partition function or field 
correlators) obtained there are insensitive to the properties 
of the ensemble of strings as a whole and depend actually on a single 
string 
only~\footnote{An obvious generalization of the string representation 
for the partition function of the theory with a global $U(1)$-symmetry 
to an ensemble of noninteracting strings 
has been performed in Ref.~\cite{lee}.}. Apart from that, there exist  
some publications 
where vortices in the 2D- and 3D Ginzburg-Landau 
theory have been treated as a grand canonical ensemble of dipoles~\cite{shap, 
bard}. The results of these investigations are as follows. 
In the 2D case~\cite{shap}, the effective field theory, 
emerging after the summation over the vortex dipoles, 
is the sine-Gordon one with the action of the 
type~\footnote{Throughout the present paper, we work in the Euclidean 
space-time.}

\begin{equation}
\label{1}
S_{\rm G-L}^{\rm 2D}=
\int d^2x\left[\left(\partial_\mu\chi\right)^2+\bar m^2\chi^2-
2\bar \zeta\cos\chi\right].
\end{equation}
Here, $\bar m$ stands for the gauge boson mass, and $\bar \zeta$ 
is the so-called fugacity of 
dimension $({\rm mass})^2$, which is the statistical weight 
(Boltzmann factor) of a single vortex dipole.
As far as the 3D case~\cite{bard} is concerned, 
there the resulting effective field 
theory 
is again the Ginzburg-Landau type theory of magnetic Higgs field 
(albeit with an additional mass 
term of the dual vector field ${\bf h}$), whose partition function 
has the form

\begin{equation}
\label{aux}
{\cal Z}_{\rm G-L}^{\rm 3D}=\int {\cal D}{\bf h}{\cal D}\Phi
{\cal D}\Phi^{*} \exp\left\{-\int d^3x\left[\frac{1}{4\eta^2}H_{\mu\nu}^2+
\frac{q^2}{2}{\bf h}^2+\left|\left(\partial_\mu+2\pi i h_\mu\right)\Phi
\right|^2+m_H^2\left|\Phi\right|^2+\lambda\left|\Phi\right|^4\right]\right\}.
\end{equation}
Here, $H_{\mu\nu}=\partial_\mu h_\nu-\partial_\nu h_\mu$ is the 
field strength tensor of the dual field ${\bf h}$, $\eta$ is the 
{\it v.e.v.} of the original Higgs field, which describes 
electric Cooper pairs, 
whose charge $q$ is the double electron one. 
In order to arrive at Eq.~(\ref{aux}), one should sum up over the 
grand canonical ensemble of the vortex dipoles, 
specifying the path-integral measure to be the one 
of the gas with a short-range repulsion as follows:

$$
{\cal Z}_{\rm G-L}^{\rm 3D}=\int {\cal D}{\bf h}
\exp\left\{-\int d^3x\left[\frac{1}{4\eta^2}H_{\mu\nu}^2+
\frac{q^2}{2}{\bf h}^2\right]\right\}\times
$$

$$
\times\left\{
1+\sum\limits_{N=1}^{\infty}\frac{1}{N!}
\left[\prod\limits_{n=1}^{N}\int\limits_{0}^{+\infty}\frac{ds_n}{s_n}
{\rm e}^{-m_H^2 s_n}
\int\limits_{{\bf x}(0)={\bf x}(s_n)}^{}{\cal D}{\bf x}\left(s_n'
\right)\right]\times\right.$$

\begin{equation}
\label{2}
\left.\times\exp\left[\sum\limits_{l=1}^{N}\int\limits_{0}^{s_l}
ds_l'\left(-\frac14
\dot{\bf x}^2\left(s_l'\right)+2\pi i\dot x_\mu\left(s_l'\right)h_\mu
\left({\bf x}\left(s_l'\right)\right)\right)
-\lambda\sum\limits_{l,k=1}^{N}
\int\limits_{0}^{s_l}ds_l'\int\limits_{0}^{s_k}ds_k''\delta\left[
{\bf x}\left(s_l'\right)-{\bf x}\left(s_k''\right)\right]\right]\right\}.
\end{equation}

Although both the free part of the 
world-line action standing in the exponent on the R.H.S. 
of Eq.~(\ref{2}) and the $\delta$-type interaction 
are quite natural, it looks desirable to perform the summation over the 
grand canonical ensemble of the vortex dipoles with the most general 
rotation- and translation invariant 
integration measure without specification of its 
concrete form. Clearly, such an approach would then enable one 
to perform also the summation over the grand canonical ensemble of 
the small vortex loops, built out of the 
Nielsen-Olesen strings in the Abelian Higgs model. 
Notice that the summation over the vortex loops seems to be  
difficult to perform by a direct generalization of Eq.~(\ref{2}), 
since the string world-sheet is a 2D object, rather than a 1D vortex 
line. 
It is our aim in the present paper to proceed with such a summation over  
the small vortex loops in the Abelian Higgs model. 
As far as 
the summation over the vortex dipoles in the Ginzburg-Landau theory
is concerned, it turns out to be 
analogous to 
the summation over the grand canonical ensemble of magnetic loops 
emerging 
in the Abelian-projected $SU(2)$-gluodynamics~\cite{su2}.
In both cases of the Ginzburg-Landau theory and Abelian Higgs model, 
the effective field theories resulting from the summation occur 
to be of the sine-Gordon type~(\ref{1}). They allow for the  
representations directly in terms of 
the integrals over the vortex dipoles (loops). As 
we shall see, besides the Biot-Savart type interaction of the objects under 
study, 
the effective actions, corresponding to these representations, 
contain a multivalued potential resulting from the cosine interaction. 
In the low-energy limit, the real branch of this  
potential takes a simple quadratic form, 
and the integration over the vortex dipoles (loops) 
becomes simply Gaussian. This will enable us to calculate exactly 
the bilocal correlation functions of the vortex dipoles and 
small vortex loops in this limit. 
After that, we shall investigate the effects brought about by such a 
way of summation over the grand canonical ensemble of the vortex dipoles 
(loops) to the Ginzburg-Landau 
theory (Abelian Higgs model) with external monopoles. 
In particular, it will be demonstrated that to the leading order 
in the inverse UV momentum cutoff, 
this method of summation 
leads to an equivalence 
of the so-extended Ginzburg-Landau theory (Abelian Higgs model) to the 
compact QED in the corresponding dimension. 

The paper is organized as follows.
In the next Section, we start our analysis with the Ginzburg-Landau 
theory and then apply the so-developed techniques in Section 3 to 
the Abelian Higgs model. The main results of the paper are summarized 
in Conclusions. Finally 
in three Appendices, some technical details 
of the performed calculations are outlined.

\section{Grand Canonical Ensemble of the 
Vortex Dipoles in the Ginzburg-Landau Theory}

Let us consider the partition function of the 3D 
Ginzburg-Landau theory in the London limit, {\it i.e.}, the limit of 
infinitely heavy Higgs field,   

\begin{equation}
\label{3}
{\cal Z}_{\rm G-L}^{\rm 3D}=\int {\cal D}A_\mu {\cal D}\theta^{\rm sing.}
{\cal D}\theta^{\rm reg.}\exp\left\{-\int d^3x\left[
\frac{1}{4q^2}F_{\mu\nu}^2+\frac{\eta^2}{2}\left(\partial_\mu\theta-
A_\mu\right)^2\right]\right\},
\end{equation}
where from now on (both in the case of the Ginzburg-Landau theory and 
Abelian Higgs model) 
we adopt the notations of Refs.~\cite{lee, orl, max, 
correl, axion, plb, correl2, epjnew}. Here, $F_{\mu\nu}=\partial_\mu A_\nu-
\partial_\nu A_\mu$ is the field strength tensor of the vector 
potential $A_\mu$, and 
$\theta=\theta^{\rm sing.}+\theta^{\rm reg.}$ stands for the 
phase of the Higgs field, consisting of a singular and a regular parts. 
First of them describes a certain configuration of Abrikosov  
vortices ({\it e.g.}, a single vortex)
according to the equation

\begin{equation}
\label{teta}
\varepsilon_{\mu\nu\lambda}\partial_\nu\partial_\lambda
\theta^{\rm sing.}({\bf x})=2\pi\delta_\mu({\bf x}).
\end{equation}
Here, $\delta_\mu({\bf x})\equiv\oint dx_\mu(\tau)\delta({\bf x}-
{\bf x}(\tau))$  
is the $\delta$-function defined {\it w.r.t.} 
the vortex line parametrized by the vector ${\bf x}(\tau)$, 
$0\le\tau\le 1$. The vortex lines are closed,
which is expressed by the equation 
$\partial_\mu\delta_\mu=0$. Notice that Eq.~(\ref{teta}) is nothing else 
but the local form of the Stokes theorem. 
As far as the usual regular part of the 
phase of the Higgs field is concerned, it describes 
fluctuations above this vortex background.

One can perform the so-called path-integral duality 
transformation (see the above mentioned Refs.), which eventually 
casts the partition 
function~(\ref{3}) into the form of the integrals over the 
dual magnetic field ${\bf h}$ and vortex lines. Referring the 
reader for details to the above cited papers, we present here 
the final result of this transformation, which has the form 

$$
{\cal Z}_{\rm G-L}^{\rm 3D}=\int {\cal D}{\bf x}(\tau){\cal D}{\bf h}
{\cal D}\varphi\exp\left\{-\int d^3x\left[\frac{1}{4\eta^2}H_{\mu\nu}^2+
\left(\frac{q}{\sqrt{2}}h_\mu+\partial_\mu\varphi\right)^2
-2\pi ih_\mu\delta_\mu\right]\right\}.
$$
The scalar field $\varphi$ can be further eliminated by performing the gauge 
transformation, $h_\mu\to h_\mu-\frac{\sqrt{2}}{q}\partial_\mu\varphi$, 
after which the partition function takes the form

\begin{equation}
\label{4}
{\cal Z}_{\rm G-L}^{\rm 3D}=\int {\cal D}{\bf x}(\tau){\cal D}{\bf h}
\exp\left\{-\int d^3x\left[\frac{1}{4\eta^2}H_{\mu\nu}^2+
\frac{q^2}{2}{\bf h}^2-2\pi ih_\mu\delta_\mu\right]\right\}.
\end{equation}
Notice that all the 
manipulations with Eq.~(\ref{4}), which will be performed  
below, should obviously be done at the 
saddle-point for the field ${\bf h}$, where 
due to the closeness of vortices, 
the second Proca equation, $\partial_\mu h_\mu=0$, 
holds.

One can now sum up over the vortex dipoles in 
Eq.~(\ref{4}) in the sense of Eq.~(\ref{2}), which leads to 
Eq.~(\ref{aux}). Our aim here is, however, 
to perform such a summation 
not specifying the measure to the 
particular form of the R.H.S. of Eq.~(\ref{2}), but rather keeping it as the 
most general rotation- and translation invariant one. To start with,     
we shall consider partition function~(\ref{4}) as a 
contribution of $N$ vortex dipoles 
to their grand canonical ensemble. 
This can be done by replacing 
$\delta_\mu$ in Eq.~(\ref{4}) by the following expression describing 
the density of the gas of the vortex dipoles 

\begin{equation}
\label{5}
\delta_\mu^{\rm gas}({\bf x})=\sum\limits_{a=1}^{N}n_a\oint 
dz_\mu^a(\tau)\delta({\bf x}-{\bf x}^a(\tau)).
\end{equation}
Here, $n_a$'s stand for winding numbers, and we have decomposed 
the vector ${\bf x}^a(\tau)$ as follows, ${\bf x}^a(\tau)={\bf y}^a+
{\bf z}^a(\tau)$, where ${\bf y}^a=\int\limits_{0}^{1}d\tau
{\bf x}^a(\tau)$ denotes the position of the $a$-th vortex dipole.
In what follows, we shall restrict ourselves to the 
minimal values of winding numbers, $n_a=\pm 1$, which is just the 
essence of the dipole approximation. That is because 
the energy of a single vortex is known to be a quadratic function 
of the flux~\cite{abrikosov}, due to which the existence 
of a dipole made out of two vortices of a unit flux is more energetically 
favorable than the existence of one vortex of the double flux. 
Besides that, we shall work in the approximation of a dilute gas of the 
vortex dipoles.  
According to this approximation, characteristic sizes of the vortex dipoles,  
$\int\limits_{0}^{1}d\tau\sqrt{\dot{\bf z}^2}$, which we shall denote 
by $a$, are much smaller than 
characteristic distances $|{\bf y}|$ between them, 
which we shall denote by $L$. In particular, this means that the vortex 
dipoles are short living objects. 

Within these two approximations, by 
substituting Eq.~(\ref{5}) into Eq.~(\ref{4}) one can proceed with 
the summation over the grand canonical ensemble of the vortex dipoles. 
This procedure 
essentially parallels a similar one of 
Ref.~\cite{su2}, and its details are outlined 
in Appendix A. As a result, we arrive at the following expression 
for the grand canonical partition function

\begin{equation}
\label{6}
{\cal Z}_{\rm grand}^{\rm 3D}=\int 
{\cal D}{\bf h}
\exp\left\{-\int d^3x\left[\frac{1}{4\eta^2}H_{\mu\nu}^2+
\frac{q^2}{2}{\bf h}^2-2\zeta\cos\left(\frac{|{\bf h}|}{\Lambda}
\right)\right]\right\}.
\end{equation}
Here, $\zeta\propto {\rm e}^{-S_0}$ is the so-called fugacity, which 
has the dimension $({\rm mass})^3$ with $S_0$ standing for the action   
of a single vortex dipole, and $\Lambda$ is the UV momentum cutoff. Thus, the 
summation over the grand canonical ensemble of the dipoles, built out of 
Abrikosov vortices, with the most general 
form of the measure of integration over their shapes yields in the 
dilute gas approximation the 
effective sine-Gordon theory~(\ref{6}). In particular, this way of 
treating the gas of the vortex dipoles 
leads to increasing the mass of the dual 
field ${\bf h}$. Namely, expanding the cosine in Eq.~(\ref{6}) we 
get the square of the full mass, $M^2=m^2+m_D^2\equiv Q^2\eta^2$, 
where $m=q\eta$ is 
the usual mass of the dual field 
(equal to the mass of the gauge boson), 
and $m_D=\frac{\eta}{\Lambda}\sqrt{2
\zeta}$ is the additional contribution coming  
from the Debye screening. We have also introduced the full electric 
charge $Q=\sqrt{q^2+\frac{2\zeta}{\Lambda^2}}$.  

Our next aim is to derive the representation 
of the partition function~(\ref{6}) directly 
in the form of an integral over the vortex dipoles. 
This can be done by making 
use of the following equality (valid at the saddle-point of the 
field ${\bf h}$, corresponding to the action standing in the 
exponent on the R.H.S. of Eq.~(\ref{4})),

$$
\exp\left\{-\int d^3x\left[\frac{1}{4\eta^2}H_{\mu\nu}^2+
\frac{q^2}{2}{\bf h}^2\right]\right\}=$$

$$
=\int {\cal D}{\bf j}\exp\left\{-\left[\frac{\pi\eta^2}{2}
\int d^3xd^3yj_\mu({\bf x})\frac{{\rm e}^{-m|{\bf x}-{\bf y}|}}{|{\bf x}-
{\bf y}|}j_\mu({\bf y})+2\pi i\int d^3x h_\mu j_\mu\right]\right\}.$$
Substituting it into Eq.~(\ref{6}), one can straightforwardly resolve 
the resulting saddle-point equation for the field ${\bf h}$, 
$\frac{h_\mu}{|{\bf h}|}\sin\left(\frac{|{\bf h}|}{\Lambda}\right)=
-\frac{i\pi\Lambda}{\zeta}j_\mu$, which yields the desired 
representation in terms of the vortex dipoles 

\begin{equation}
\label{7}
{\cal Z}_{\rm grand}^{\rm 3D}=
\int {\cal D}{\bf j}\exp\left\{-\left[\frac{\pi\eta^2}{2}
\int d^3xd^3yj_\mu({\bf x})\frac{{\rm e}^{-m|{\bf x}-{\bf y}|}}{|{\bf x}-
{\bf y}|}j_\mu({\bf y})+V[2\pi {\bf j}]\right]\right\}.
\end{equation}
Here, the complex-valued potential of the vortex dipoles reads 
({\it cf.} Ref.~\cite{su2}) 

\begin{equation}
\label{8}
V[{\bf j}]=\sum\limits_{n=-\infty}^{+\infty}
\int d^3x\left\{\Lambda |{\bf j}|\left[\ln\left[
\frac{\Lambda}{2\zeta}|{\bf j}|+\sqrt{1+\left(
\frac{\Lambda}{2\zeta}|{\bf j}|\right)^2}\right]+2\pi in\right]- 2\zeta
\sqrt{1+\left(
\frac{\Lambda}{2\zeta}|{\bf j}|\right)^2}
\right\}.
\end{equation}

The obtained representation~(\ref{7}) can now be applied 
to the calculation of correlators of the vortex dipoles. 
Indeed, it is possible to  
demonstrate that if we introduce into Eq.~(\ref{4}) (with $\delta_\mu$ 
replaced by $\delta_\mu^{\rm gas}$) a unity of the form 

$$1=\int {\cal D}{\bf j}\delta\left(j_\mu-
\delta_\mu^{\rm gas}
\right)=\int {\cal D}{\bf j}{\cal D}{\bf l}\exp\left[
-2\pi i\int d^3x l_\mu\left(j_\mu-\delta_\mu^{\rm gas}\right)
\right]$$
and integrate out all the fields except ${\bf j}$, the result will 
coincide with Eq.~(\ref{7}). This is the reason why the correlators 
of ${\bf j}$'s are nothing else, but the correlators of the vortex dipoles.
Such correlators can be calculated in the low-energy limit, {\it i.e.},
when $\Lambda|{\bf j}|\ll\zeta$ and one considers in Eq.~(\ref{8}) the 
stable minimum of the real branch of the potential by extracting 
from the whole sum the term with $n=0$. In this case, the bilocal 
correlator of the vortex dipoles reads 

$$\left<j_\mu({\bf z})j_\nu({\bf 0})\right>\equiv\frac{
\int {\cal D}{\bf j} j_\mu({\bf z})j_\nu({\bf 0})
\exp\left\{-\left[\frac{\pi\eta^2}{2}
\int d^3xd^3yj_\mu({\bf x})
\frac{{\rm e}^{-m|{\bf x}-{\bf y}|}}{|{\bf x}-
{\bf y}|}j_\mu({\bf y})+\int d^3x\left(-2\zeta+\frac{\pi^2\Lambda^2
{\bf j}^2}{\zeta}\right)\right]\right\}} 
{\int {\cal D}{\bf j} 
\exp\left\{-\left[\frac{\pi\eta^2}{2}
\int d^3xd^3yj_\mu({\bf x})
\frac{{\rm e}^{-m|{\bf x}-{\bf y}|}}{|{\bf x}-
{\bf y}|}j_\mu({\bf y})+\int d^3x\left(-2\zeta+\frac{\pi^2\Lambda^2
{\bf j}^2}{\zeta}\right)\right]\right\}}=$$

\begin{equation}
\label{biloc}
=\delta_{\mu\nu}\frac{m_D^2}{4\pi^2\eta^2}\left[\delta({\bf z})-
m_D^2\frac{1}{4\pi}\frac{{\rm e}^{-M|{\bf z}|}}{|{\bf z}|}\right].
\end{equation}
This example illustrates how the vortex dipoles in the 
grand canonical ensemble are correlated to each other. In particular, 
we see that their correlator decreases exactly according to the Yukawa 
law with the screening provided by the full mass $M$. 

Let us now see what will be the consequences of accounting for interaction 
of the vortex dipoles in the grand canonical ensemble if we introduce into the 
Ginzburg-Landau theory external monopoles. 
To start with, we shall introduce the monopoles into the 
Ginzburg-Landau theory with noninteracting Abrikosov vortices. 
This can be done by replacing 
the field strength tensor $F_{\mu\nu}$ in Eq.~(\ref{3}) by 
$F_{\mu\nu}+F_{\mu\nu}^M$, where the monopole field strength tensor 
$F_{\mu\nu}^M$ obeys the equation $\frac12\varepsilon_{\mu\nu\lambda}
\partial_\mu F_{\nu\lambda}^M=2\pi\rho$ with $\rho$ standing for the 
density of monopoles. The path-integral duality transformation of 
the so-extended theory~(\ref{3}) has been performed in Ref.~\cite{epjnew} 
and effectively results to adding to the Lagrangian standing in the 
exponent on the R.H.S. of Eq.~(\ref{4}) the term 

\begin{equation}
\label{9}
\frac{i}{2}h_\mu
\frac{\partial}{\partial x_\mu}\int d^3y\frac{\rho({\bf y})}{|{\bf x}-
{\bf y}|}.  
\end{equation}

We are now in the position to 
investigate what will be the effect of summation over the grand canonical 
ensemble of interacting vortex dipoles to the Ginzburg-Landau theory with 
external monopoles.
If we restrict ourselves to the leading term in the 
$\frac{1}{\Lambda}$-expansion of Eq.~(\ref{6}),  
{\it i.e.}, keep in the expansion of the cosine only the term 
quadratic in ${\bf h}$, than the integration over ${\bf h}$ is 
Gaussian, and the result has the form 

\begin{equation}
\label{11}
{\cal Z}_{\rm grand}^{\rm 3D}[\rho]
\stackrel{{\Lambda\to\infty}}{\longrightarrow} 
\exp\left[-\frac{\pi}{2Q^2}\int d^3xd^3y\frac{\rho({\bf x})
\rho({\bf y})}{|{\bf x}-{\bf y}|}\right].
\end{equation}
The details of a derivation of Eq.~(\ref{11}) are presented in Appendix B.  
This equation means 
that in the physical limit, when the UV momentum cutoff 
infinitely increases,  
the partition function of the Ginzburg-Landau theory 
with external monopoles, where the vortex dipoles are summed up 
in the sense of the grand canonical ensemble, is  
equivalent to the statistical weight of 
3D compact QED with the charge $q$ of Cooper pairs 
replaced by the full one $Q$.

Finally, it is worth noting that 
accounting for all the terms 
in the 
$\frac{1}{\Lambda}$-expansion of Eq.~(\ref{6}), 
rather than only for the leading one, 
leads simply to the following substitution in Eq.~(\ref{7})

$$V[2\pi {\bf j}]\longrightarrow V\left[2\pi{\bf j}+
\frac12\nabla_{{\bf x}}
\int d^3y\frac{\rho({\bf y})}{|{\bf x}-{\bf y}|}\right].$$

\section{Grand Canonical Ensemble of the Small Vortex Loops 
in the Abelian Higgs Model}

In the present Section, we shall extend 
the analysis of the grand canonical 
ensemble of the vortex dipoles to the 4D case of the small vortex loops, 
built out of Nielsen-Olesen strings,  
in the London limit of the Abelian Higgs model. 
The partition function under study is given by Eq.~(\ref{3}) with the 
replacement $d^3x\to d^4x$. Notice that in what follows, we shall 
mark the 4D quantities ({\it e.g.}, charge of Cooper pairs, {\it v.e.v.} 
of the Higgs field, {\it etc.}) by prime in order to distinguish them 
from the corresponding 3D ones. 
Equation~(\ref{teta}) 
goes over into~\cite{lee, orl, max, 
correl, axion, epjnew}  

$$
\varepsilon_{\mu\nu\lambda\rho}\partial_\lambda\partial_\rho
\theta^{\rm sing.}(x)=2\pi\Sigma_{\mu\nu}(x),$$
where $\Sigma_{\mu\nu}(x)\equiv\int\limits_{\Sigma}^{}d\sigma_{\mu\nu}
(x(\xi))\delta(x-x(\xi))$ is the so-called vorticity tensor current  
defined at the string world-sheet $\Sigma$, parametrized 
by the vector $x_\mu(\xi)$. Here, $\xi=\left(\xi^1,\xi^2\right)\in 
[0,1]\times [0,1]$ denotes 
the 2D coordinate, and $d\sigma_{\mu\nu}(x(\xi))$ stands for the  
infinitesimal world-sheet element. 
Due to the closeness of strings, the vorticity 
tensor current is conserved, {\it i.e.}, $\partial_\mu\Sigma_{\mu\nu}=0$. 

The path-integral duality transformation of the Abelian Higgs model 
accounting for noninteracting 
Nielsen-Olesen strings has been performed in the above 
mentioned papers, and the result reads 

$$
{\cal Z}_{\rm AHM}=
$$

$$
=\int {\cal D}x_\mu(\xi){\cal D}h_{\mu\nu}{\cal D}B_\mu 
\exp\left\{-\int d^4x\left[\frac{1}{12\eta'^2}H_{\mu\nu\lambda}^2+
\left(\frac{q'}{2}h_{\mu\nu}+\partial_\mu B_\nu-\partial_\nu B_\mu
\right)^2-i\pi h_{\mu\nu}\Sigma_{\mu\nu}\right]
\right\}.
$$
Here, $H_{\mu\nu\lambda}\equiv\partial_\mu h_{\nu\lambda}+
\partial_\lambda h_{\mu\nu}+\partial_\nu h_{\lambda\mu}$ stands for the 
field strength tensor of the massive antisymmetric tensor field 
$h_{\mu\nu}$, usually referred to as the Kalb-Ramond field~\cite{kalb}.
Analogously to the 3D case, the vector field $B_\mu$ can be eliminated 
by performing the hypergauge transformation, $h_{\mu\nu}\to h_{\mu\nu}-
\frac{2}{q'}\left(\partial_\mu B_\nu-\partial_\nu B_\mu\right)$, after which 
the partition function takes the form 

\begin{equation}
\label{12}
{\cal Z}_{\rm AHM}=\int {\cal D}x_\mu(\xi){\cal D}h_{\mu\nu}
\exp\left\{-\int d^4x\left[\frac{1}{12\eta'^2}H_{\mu\nu\lambda}^2+
\frac{q'^2}{4}h_{\mu\nu}^2-i\pi h_{\mu\nu}\Sigma_{\mu\nu}\right]
\right\}.
\end{equation}
In order to proceed from the individual string to the gas 
of the small vortex loops, 
one should analogously to the 3D case substitute 
for $\Sigma_{\mu\nu}$ in Eq.~(\ref{12}) the following expression  

\begin{equation}
\label{13}
\Sigma_{\mu\nu}^{\rm gas}(x)=\sum\limits_{a=1}^{N}n_a\int 
d\sigma_{\mu\nu}\left(x^a(\xi)\right)\delta\left(x-x^a(\xi)\right).
\end{equation}
After that, the summation over the 
grand canonical ensemble of the vortex loops is straightforward.
Referring the reader for the details to 
Appendix C, we shall present here the result of this procedure, 
which has the form

\begin{equation}
\label{14}
{\cal Z}_{\rm grand}^{\rm 4D}=\int {\cal D}h_{\mu\nu}\exp
\left\{-\int d^4x\left[\frac{1}{12\eta'^2}H_{\mu\nu\lambda}^2+
\frac{q'^2}{4}h_{\mu\nu}^2-2\zeta'\cos\left(\frac{\left|h_{\mu\nu}
\right|}{\Lambda'^2}\right)\right]\right\}.
\end{equation} 
Here, $\left|h_{\mu\nu}\right|\equiv\sqrt{h_{\mu\nu}^2}$, $\Lambda'$ 
is a new UV momentum cutoff, and 
the fugacity $\zeta'$ (Boltzmann factor of a single vortex loop) 
has now the dimension $({\rm mass})^4$. The square of the 
full mass of the field $h_{\mu\nu}$ following from Eq.~(\ref{14}) 
reads $M'^2=m'^2+m_D'^2\equiv Q'^2\eta'^2$. Here,  
$m'=q'\eta'$ is the usual Higgs contribution, 
$m_D'=\frac{2\eta'\sqrt{\zeta'}}{\Lambda'^2}$ is the Debye contribution, and 
$Q'=\sqrt{q'^2+\frac{4\zeta'}{\Lambda'^4}}$ is the full electric charge. 

The representation of the partition function~(\ref{14}) in terms of the 
vortex loops can be obtained 
by virtue of the following equality 
valid at the saddle-point of Eq.~(\ref{12}) (with $\Sigma_{\mu\nu}\to 
\Sigma_{\mu\nu}^{\rm gas}$), at which $\partial_\mu h_{\mu\nu}=0$,

$$
\exp
\left\{-\int d^4x\left[\frac{1}{12\eta'^2}H_{\mu\nu\lambda}^2+
\frac{q'^2}{4}h_{\mu\nu}^2\right]\right\}=$$

$$
=\int {\cal D}S_{\mu\nu}\exp\left\{-\left[\frac{q'\eta'^3}{4}
\int d^4xd^4y S_{\mu\nu}(x)\frac{K_1\left(m'|x-y|\right)}{|x-y|}
S_{\mu\nu}(y)+i\pi\int d^4x h_{\mu\nu}S_{\mu\nu}\right]\right\},$$
where $K_1$ stands for the modified Bessel function. Substituting 
this equality into Eq.~(\ref{14}), we can integrate the field 
$h_{\mu\nu}$ out, which yields the desired representation 
for the partition function~(\ref{14}), 

\begin{equation}
\label{15}
{\cal Z}_{\rm grand}^{\rm 4D}=\int {\cal D}S_{\mu\nu}\exp\left\{
-\left[\frac{q'\eta'^3}{4} 
\int d^4xd^4y S_{\mu\nu}(x)\frac{K_1\left(m'|x-y|\right)}{|x-y|}
S_{\mu\nu}(y)+V\left[\pi\Lambda'S_{\mu\nu}\right]\right]\right\},
\end{equation}
where the effective potential $V$ is given by Eq.~(\ref{8}) 
with $d^3x\to d^4x$. 

Similarly to the 3D case, correlation functions of $S_{\mu\nu}$'s, 
calculated by virtue of the partition function~(\ref{15}), are nothing 
else, but the correlation functions of the small vortex loops 
in the gas. This can 
be seen by mentioning that if we insert into Eq.~(\ref{12}) with 
$\Sigma_{\mu\nu}$ replaced by $\Sigma_{\mu\nu}^{\rm gas}$ the following  
unity 

$$1=\int {\cal D}S_{\mu\nu}\delta\left(S_{\mu\nu}-
\Sigma_{\mu\nu}^{\rm gas}\right)=\int {\cal D}S_{\mu\nu}{\cal D}
l_{\mu\nu}\exp\left[-i\pi\int d^4x l_{\mu\nu}\left(S_{\mu\nu}-
\Sigma_{\mu\nu}^{\rm gas}\right)\right]$$
and integrate out all the fields except $S_{\mu\nu}$, the result 
will coincide with Eq.~(\ref{15}). Such correlation functions of the small 
vortex loops    
can be most easily calculated in the low-energy limit, $\Lambda'^2
\left|S_{\mu\nu}\right|\ll\zeta'$, by considering the stable minimum 
of the real branch of the potential~(\ref{8}), where it takes a simple 
parabolic form. In particular, the bilocal correlation function  
reads ({\it cf.} Eq.~(\ref{biloc}))

$$
\left<S_{\alpha\beta}(z)S_{\lambda\rho}(0)\right>\equiv$$

$$\equiv
\frac{
\int {\cal D}S_{\mu\nu} S_{\alpha\beta}(z)S_{\lambda\rho}(0)
\exp\left\{
-\left[\frac{q'\eta'^3}{4} 
\int d^4xd^4y S_{\mu\nu}(x)\frac{K_1\left(m'|x-y|\right)}{|x-y|}
S_{\mu\nu}(y)+\frac{\pi^2\Lambda'^4}{4\zeta'}\int d^4x S_{\mu\nu}^2
\right]\right\}}{\int {\cal D}S_{\mu\nu}
\exp\left\{
-\left[\frac{q'\eta'^3}{4} 
\int d^4xd^4y S_{\mu\nu}(x)\frac{K_1\left(m'|x-y|\right)}{|x-y|}
S_{\mu\nu}(y)+\frac{\pi^2\Lambda'^4}{4\zeta'}\int d^4x S_{\mu\nu}^2
\right]\right\}}=$$

$$=\left(\delta_{\alpha\lambda}\delta_{\beta\rho}-\delta_{\alpha\rho}
\delta_{\beta\lambda}\right)\frac{m_D'^2}{4\pi^2\eta'^2}\left[\delta(z)-
m_D'^2\frac{M'}{4\pi^2}\frac{K_1\left(M'|z|\right)}{|z|}\right].$$
Thus, we see that if Nielsen-Olesen strings are considered not as 
individual ones, but as a gas of the small vortex loops, 
the interaction between which has the 
Yukawa form, the bilocal correlator of the vortex loops 
in the low-energy limit  
has also the Yukawa behaviour 
(albeit screened not by the mass of the gauge boson, but by the full 
mass $M'$). 

Let us now investigate grand canonical ensemble of the vortex loops 
in the presence of external monopoles. To start with, we first clarify 
what will be the modifications of the Abelian Higgs model with  
Nielsen-Olesen strings treated as individual ones 
due to external monopoles. 
The starting partition function has the form 

\begin{equation}
\label{16}
{\cal Z}_{\rm AHM}\left[j_\mu\right]=
\int {\cal D}A_\mu {\cal D}\theta^{\rm sing.}
{\cal D}\theta^{\rm reg.}\exp\left\{-\int d^4x\left[
\frac{1}{4q'^2}\left(F_{\mu\nu}+F_{\mu\nu}^{M'}\right)^2+
\frac{\eta'^2}{2}\left(\partial_\mu\theta-
A_\mu\right)^2\right]\right\}.
\end{equation}
Here, the monopole field strength tensor $F_{\mu\nu}^{M'}$ obeys the 
equation $\partial_\mu\tilde F_{\mu\nu}^{M'}=j_\nu$, where $\tilde 
F_{\mu\nu}^{M'}\equiv\frac12\varepsilon_{\mu\nu\lambda\rho}
F_{\lambda\rho}^{M'}$, 
according to which

\begin{equation}
\label{17}
F_{\mu\nu}^{M'}(x)=-\frac{1}{4\pi^2}\varepsilon_{\mu\nu\lambda\rho}
\frac{\partial}{\partial x_\lambda} 
\int d^4y\frac{j_\rho(y)}{(x-y)^2}
\end{equation}
with $j_\mu$ standing for the (conserved) monopole current.
The path-integral duality 
transformation of the partition function~(\ref{16}) has been performed 
in Ref.~\cite{correl2} and leads to Eq.~(\ref{12}) with the substitution 
$\Sigma_{\mu\nu}\to\hat\Sigma_{\mu\nu}\equiv
\Sigma_{\mu\nu}+\frac{1}{2\pi}\tilde F_{\mu\nu}^{M'}$.

It is now straightforward to see what will be the consequences of the 
summation over the grand canonical ensemble of the small vortex loops 
to the Abelian 
Higgs model with external monopoles.
Namely, this summation leads to the following expression for the 
partition function: 

\begin{equation}
\label{21}
{\cal Z}_{\rm grand}^{\rm 4D}\left[j_\mu\right]=\int {\cal D}h_{\mu\nu}
\exp\left\{-\int d^4x\left[\frac{1}{12\eta'^2}H_{\mu\nu\lambda}^2+
\frac{q'^2}{4}h_{\mu\nu}^2-2\zeta'\cos\left(\frac{\left|h_{\mu\nu}
\right|}{\Lambda'^2}\right)-\frac{i}{2}h_{\mu\nu}\tilde F_{\mu\nu}^{M'} 
\right]\right\}. 
\end{equation}
In particular, in the physical limit $\Lambda'\to\infty$, keeping 
in Eq.~(\ref{21}) only the leading term in the 
$\frac{1}{\Lambda'}$-expansion, we obtain 

\begin{equation}
\label{aux1}
{\cal Z}_{\rm grand}^{\rm 4D}
\left[j_\mu\right]\stackrel{{\Lambda'\to\infty}}{
\longrightarrow}\exp\left[-\frac{1}{8\pi^2Q'^2}\int d^4xd^4y
j_\mu(x)\frac{1}{(x-y)^2}j_\mu(y)\right].
\end{equation}
The derivation of Eq.~(\ref{aux1}) is similar to that of Eq.~(\ref{11}). 
The only 
technical detail necessary for it is the integration over the Kalb-Ramond 
field, which can be found in Refs.~\cite{plb1, correl2}. 

Thus we see that similarly to the 3D case, in this physical limit, the 
partition function of the grand canonical ensemble 
of the small vortex loops extended 
by external monopoles is equivalent to the statistical weight 
of the 4D compact QED with 
the electric charge $q'$ replaced by the full one $Q'$. 

Finally, accounting for all the orders in the $1/\Lambda'$-expansion 
in Eq.~(\ref{21}) is equivalent to the following substitution  
in Eq.~(\ref{15}): 

$$
V\left[\pi\Lambda'S_{\mu\nu}\right]\longrightarrow 
V\left[\Lambda'\left(\pi S_{\mu\nu}-\frac12\tilde F_{\mu\nu}^{M'} 
\right)\right].$$
This means that external monopoles do not affect the Yukawa-type 
interaction of the vortex loops, but enter only their effective potential.

\section{Conclusions}

In the present paper, we have investigated grand canonical 
ensembles of the vortex dipoles and 
small vortex loops in the London limit of the Ginzburg-Landau theory and 
Abelian Higgs model, respectively. In the approximation where 
these objects form a dilute gas, the summation over them 
with the most general rotation- and translation 
invariant measure of integration over their shapes has been performed.
The resulting effective theories turned out to have the form of the 
sine-Gordon theories due to the additional 
cosine interaction term of the dual field, emerging from the above mentioned 
summation. In the physical limit of the large UV momentum cutoff, one can 
keep in the expansion of this cosine interaction  
in powers of the inverse cutoff only the first term. This yields a certain  
(positive) correction to the mass of the dual field due to the 
Debye screening. 

After that, we have casted the obtained sine-Gordon theories, corresponding 
to the Ginzburg-Landau theory and Abelian Higgs model, into the forms 
of the representations in terms of the vortex dipoles and small vortex 
loops, respectively. 
The resulting effective actions turned out to have 
similar forms and consist of the Biot-Savart Yukawa type interaction of 
the objects under study and 
a certain multivalued effective potential. In the 
low-energy limit, this potential takes a simple quadratic form, and 
the effective actions become Gaussian. This enables one to calculate 
correlation functions of the vortex dipoles (loops) 
in this limit, and as an example, 
we have calculated the bilocal correlators. Those turned out to have 
the Yukawa form, where the screening is governed  
by the full mass of the dual field, 
resulting both from the Higgs and Debye effects. 

Then, we have addressed the problem of what will be the consequences 
of our approach to treatment the vortex dipoles (loops) as a grand 
canonical ensemble to the theories under study extended by 
external monopoles. In this way, we have demonstrated that to the 
leading order in the expansion in powers of the inverse momentum cutoff, 
summation over the grand canonical ensemble results to the complete 
equivalence of these theories to the compact QED in the 
corresponding dimension with the charge of  
Cooper pairs 
replaced by the full one (which accounts also for the Debye screening). 
This result differs from that, which one gets by considering Abrikosov 
vortices (Nielsen-Olesen strings) 
as individual ones. In that case, the equivalence to compact 
QED holds only in the limit of vanishing gauge boson mass. 
As far as the effects brought about by external monopoles to all the orders 
of the expansion in the inverse powers of the UV momentum cutoff are 
concerned, those result into certain shifts of the arguments of the 
effective potentials, but do not affect the Biot-Savart 
interaction.

It now looks attractive to investigate the grand canonical ensembles 
of the vortex dipoles (loops) 
emerging in the effective Abelian-projected theories, 
corresponding to the original $SU(N)$, $N>2$, Yang-Mills theories, 
since the dual Abelian Higgs model can be under some assumptions considered 
as such an effective theory for the case $N=2$. Recently, some progress
in this direction has been achieved in Ref.~\cite{prep}.

\section*{Acknowledgments}

The author is indebted to Profs. D. Ebert and A. Wipf for 
valuable discussions. He is also grateful to Prof. H. Kleinert and 
the theoretical physics group 
of the Freie University of Berlin, where some part of the present 
work has been done, for kind hospitality.

\section*{Appendix A. Summation over the Grand Canonical Ensemble of 
the Vortex Dipoles}

In this Appendix, we shall present some details of a derivation 
of Eq.~(\ref{6}) of the main text. 
Substituting Eq.~(\ref{5}) into Eq.~(\ref{4}), 
one can perform the summation over the grand canonical ensemble 
of the vortex dipoles as follows: 

$$
1+\sum\limits_{N=1}^{\infty}\frac{\zeta^N}{N!}\left(\prod\limits_{i=1}^{N}
\int d^3y^i\int {\cal D} {\bf z}^i\mu\left[{\bf z}^i\right]\right)
\sum\limits_{n_a=\pm 1}^{}\exp\left\{2\pi i\sum\limits_{a=1}^{N}
n_a\oint dz_\mu^a h_\mu\left({\bf x}^a\right)\right\}=$$

$$
=1+\sum\limits_{N=1}^{\infty}\frac{(2\zeta)^N}{N!}\left\{
\int d^3y \int {\cal D}{\bf z}\mu[{\bf z}]\cos\left(2\pi
\oint dz_\mu h_\mu({\bf x})\right)\right\}^N.$$
Here, we have introduced the so-called fugacity $\zeta$, which is 
proportional to the statistical weight of a single vortex dipole and 
has the dimension $({\rm mass})^3$. 
Next, $\mu[{\bf z}]$ stands for a certain rotation- and 
translation invariant measure of integration over the shapes of dipoles.
Let us now employ for the grand canonical ensemble of the vortex dipoles under 
consideration the dilute gas approximation described after Eq.~(\ref{5}). 
One can then expand $h_\mu({\bf x})$ up to the first order 
in $a/L$ as follows: 

$$
h_\mu({\bf x})=h_\mu({\bf y})+L^{-1}z_\nu n_\nu h_\mu({\bf y})+
{\cal O}\left(\left(\frac{a}{L}\right)^2\right),\eqno (A.1)
$$
where $n_\nu=y_\nu/|{\bf y}|$, and we have substituted $n_\nu/L$ for 
the derivative $\partial/\partial y_\nu$. By making use of this 
expansion, we obtain

$$
\int {\cal D}{\bf z}\mu[{\bf z}]\cos\left(2\pi\oint dz_\mu h_\mu
({\bf x})\right)\simeq \int {\cal D}{\bf z}\mu[{\bf z}]\cos\left(
\frac{2\pi}{L}n_\nu h_\mu({\bf y})P_{\mu\nu}[{\bf z}]\right)=$$

$$
=\sum\limits_{n=0}^{\infty}\frac{(-1)^n}{(2n)!}\left(\frac{2\pi}{L}
\right)^{2n}n_{\nu_1}h_{\mu_1}({\bf y})\cdots n_{\nu_{2n}}
h_{\mu_{2n}}({\bf y})\int {\cal D}{\bf z}\mu[{\bf z}]
P_{\mu_1\nu_1}[{\bf z}]\cdots P_{\mu_{2n}\nu_{2n}}[{\bf z}]=$$

$$
=\sum\limits_{n=0}^{\infty}\frac{(-1)^n}{(2n)!}\left(\frac{2\pi a^2}{L}
\right)^{2n}n_{\nu_1}h_{\mu_1}({\bf y})\cdots n_{\nu_{2n}}
h_{\mu_{2n}}({\bf y})\times$$

$$\times\frac{1}{(2n-1)!!}\left[\hat 1_{\mu_1\nu_1, 
\mu_2\nu_2}\cdots \hat 1_{\mu_{2n-1}\nu_{2n-1}, \mu_{2n}\nu_{2n}}+{\,}
{\rm permutations}\right].$$
Here, $P_{\mu\nu}[{\bf z}]\equiv\oint dz_\mu z_\nu$ is the so-called 
tensor area, and $\hat 1_{\mu\nu, \lambda\rho}=
\frac12\left(\delta_{\mu\lambda}\delta_{\nu\rho}-\delta_{\mu\rho}
\delta_{\nu\lambda}\right)$. Notice that the sum in square brackets on the 
R.H.S. of the last equality contains $(2n-1)!!$ terms, which is the 
reason for extracting explicitly this normalization factor. 
The general form of the tensor structure standing in these brackets 
is due to the rotation- and translation invariance of the measure 
$\mu[{\bf z}]$. In the further contraction of indices, it is worth 
noting that within the dilute gas approximation, 
$n_\mu h_\mu({\bf y})=0$. Indeed, as we have already seen, 
this approximation allows for the 
substitution $n_\mu\to L\frac{\partial}{\partial y_\mu}$. On the 
other hand, the divergency of the field ${\bf h}$ vanishes 
according to the second Proca equation,
valid at the saddle point of the 
field ${\bf h}$ corresponding to Eq.~(\ref{4}). This 
completes the proof of our statement.    

Finally, introducing 
the UV momentum cutoff as
$\Lambda=\frac{L}{\sqrt{2}\pi a^2}{\,}\left(\gg a^{-1}\right)$,  
we obtain 

$$
\int {\cal D}{\bf z}\mu[{\bf z}]\cos\left(2\pi\oint dz_\mu h_\mu
({\bf x})\right)\simeq\cos\left(\frac{|{\bf h}({\bf y})|}{\Lambda}
\right),$$
which leads to Eq.~(\ref{6}). 

\section*{Appendix B. Details of a Derivation of Eq.~(\ref{11})}

In the present Appendix, some details of a derivation of Eq.~(\ref{11}) 
will be presented. Let us start with the following expression: 

$$
J\equiv\int {\cal D}{\bf h}\exp\left\{-\int d^3x\left[
\frac{1}{4\eta^2}H_{\mu\nu}^2+\frac{Q^2}{2}{\bf h}^2+
\frac{i}{2}h_\mu\frac{\partial}{\partial x_\mu}\int d^3y
\frac{\rho({\bf y})}{|{\bf x}-{\bf y}|}\right]\right\},$$
which (up to an inessential constant factor, which can be referred to the 
integration measure) is the leading term in the $\frac{1}{\Lambda}$-expansion 
of the partition function~(\ref{6}) including external 
monopole part~(\ref{9}). Gaussian integration over the field ${\bf h}$ yields

$$
J=\exp\left\{-\frac{\eta^2}{2}\int d^3xd^3y\frac{{\rm e}^{-M|{\bf x}
-{\bf y}|}}{4\pi|{\bf x}-{\bf y}|}\left[\frac14\frac{\partial^2}{\partial 
x_\mu\partial y_\mu}\int d^3z\frac{\rho({\bf z})}{|{\bf x}-{\bf z}|}
\int d^3u\frac{\rho({\bf u})}{|{\bf y}-{\bf u}|}+\frac{4\pi^2}{M^2}
\rho({\bf x})\rho({\bf y})\right]\right\},\eqno (B.1)$$
where the normalization factor was assumed to be included into the measure 
${\cal D}{\bf h}$. 

Let us consider the first term in square brackets on the R.H.S. of 
Eq.~(B.1) together with the factor in front of these brackets. 
By performing the partial integration, 
one can cast the derivative $\partial/\partial y_\mu$ to the factor 
in front of the brackets, after which due to the translation invariance 
of this factor it can be replaced by $-\partial/\partial x_\mu$. 
This derivative can be casted back by doing one more partial integration, 
which yields for this group of terms 

$$
\frac14\int d^3xd^3y\frac{{\rm e}^{-M|{\bf x}
-{\bf y}|}}{4\pi|{\bf x}-{\bf y}|}\frac{\partial^2}{\partial 
x_\mu\partial y_\mu}\int d^3z\frac{\rho({\bf z})}{|{\bf x}-{\bf z}|}
\int d^3u\frac{\rho({\bf u})}{|{\bf y}-{\bf u}|}=$$

$$=-\frac14 
\int d^3xd^3y\frac{{\rm e}^{-M|{\bf x}
-{\bf y}|}}{4\pi|{\bf x}-{\bf y}|}\Delta_{\bf x}
\int d^3z\frac{\rho({\bf z})}{|{\bf x}-{\bf z}|}
\int d^3u\frac{\rho({\bf u})}{|{\bf y}-{\bf u}|}=
\pi\int d^3xd^3yd^3u\rho({\bf x})\frac{{\rm e}^{-M|{\bf x}
-{\bf y}|}}{4\pi|{\bf x}-{\bf y}|}\frac{\rho({\bf u})}{|{\bf y}-{\bf u}|}.$$
Substituting this expression into Eq.~(B.1), we obtain

$$
J=\exp\left\{-\frac{\pi\eta^2}{2}\int d^3xd^3y
\frac{{\rm e}^{-M|{\bf x}
-{\bf y}|}}{4\pi|{\bf x}-{\bf y}|}
\left[\int d^3u 
\frac{\rho({\bf x})\rho({\bf u})}{|{\bf y}-{\bf u}|}+
\frac{4\pi}{M^2}\rho({\bf x})\rho({\bf y})\right]\right\}.\eqno (B.2)$$

The integral 

$$I\equiv\int d^3y\frac{{\rm e}^{-M|{\bf x}
-{\bf y}|}}{|{\bf x}-{\bf y}||{\bf y}-{\bf u}|}=\int d^3y
\frac{{\rm e}^{-M|{\bf y}|}}{|{\bf y}||{\bf w}-{\bf y}|},\eqno (B.3)$$
where ${\bf w}\equiv {\bf x}-{\bf u}$, can be calculated by 
the two alternative ways. 
One of them is to divide the integration region over $|{\bf y}|$ into 
two parts, $[0,|{\bf w}|]$, $[|{\bf w}|,+\infty)$ and 
expand $\frac{1}{|{\bf w}-{\bf y}|}$ in 
Legendre polynomials $P_n$'s on both of them. Then, 
the integration over the azimuthal 
angle singles out from the whole series only the zeroth term, 
$\int\limits_{-1}^{1}P_n(\cos\theta)d\cos\theta=2\delta_{n0}$. After that, 
the integration over $|{\bf y}|$ at both intervals is straightforward 
and yields for the integral~(B.3) the following result:  

$$I=
\frac{4\pi}{M^2|{\bf w}|}\left(1-{\rm e}^{-M|{\bf w}|}\right).\eqno (B.4)$$
Another way to calculate the integral~(B.3) is to use the relations 

$$
\int\frac{d^3p}{(2\pi)^3}\frac{{\rm e}^{i{\bf p}{\bf y}}}{{\bf p}^2+M^2}=
\frac{1}{4\pi}\frac{{\rm e}^{-M|{\bf y}|}}{|{\bf y}|}~~ {\rm and}~~ 
\int\frac{d^3q}{(2\pi)^3}
\frac{{\rm e}^{i{\bf q}({\bf w}-{\bf y})}}{{\bf q}^2}=
\frac{1}{4\pi|{\bf w}-{\bf y}|},$$
by virtue of which we have 

$$
I=
16\pi^2\int d^3y\int\frac{d^3pd^3q}{(2\pi)^6}
\frac{{\rm e}^{i{\bf p}{\bf y}+i{\bf q}({\bf w}-{\bf y})}}{\left({\bf p}^2+
M^2\right){\bf q}^2}=
16\pi^2\int\frac{d^3p}{(2\pi)^3}\frac{{\rm e}^{i{\bf p}{\bf w}}}{{\bf p}^2
\left({\bf p}^2+M^2\right)}=$$

$$=16\pi^2\int\frac{d^3p}{(2\pi)^3}
\int\limits_{0}^{+\infty}d\alpha\int\limits_{0}^{+\infty}d\beta
{\rm e}^{i{\bf p}{\bf w}-\alpha {\bf p}^2-\beta\left({\bf p}^2+M^2\right)}=
2\sqrt{\pi}\int\limits_{0}^{+\infty}d\alpha\int\limits_{0}^{+\infty}d\beta
\frac{{\rm e}^{-\beta M^2-\frac{{\bf w}^2}{4(\alpha+\beta)}}}{(\alpha+
\beta)^{\frac32}}.$$
It is now suitable to introduce new integration variables as 
$\alpha=bt$, $\beta=b(1-t)$; $b\in [0,+\infty)$, $t\in [0,1]$, which yields 

$$
I=2\sqrt{\pi}
\int\limits_{0}^{+\infty}\frac{db}{\sqrt{b}}\int\limits_{0}^{1}dt
{\rm e}^{-\frac{{\bf w}^2}{4b}-bM^2(1-t)}=\frac{2\sqrt{\pi}}{M^2}
\int\limits_{0}^{+\infty}\frac{db}{b^{\frac32}}{\rm e}^{-\frac{{\bf w}^2}{4b}}
\left(1-{\rm e}^{-bM^2}\right).$$
The remaining integral is already straightforward to evaluate by changing 
the variable $b\to\frac{1}{b}$ and making use of the formula

$$
\int\limits_{0}^{+\infty}\frac{db}{\sqrt{b}}{\rm e}^{-Ab-\frac{B}{b}}=
\sqrt{\frac{\pi}{A}}{\rm e}^{-2\sqrt{AB}},$$
valid for positive $A$ and $B$. The result coincides with Eq.~(B.4).

Finally, substituting Eq.~(B.4) into Eq.~(B.2) and recalling that 
$M=Q\eta$, we arrive at Eq.~(\ref{11}) of the main text.

\section*{Appendix C. Summation over the Grand Canonical Ensemble of 
the Small Vortex Loops}

In the present Appendix, we shall outline some steps of a derivation 
of Eq.~(\ref{14}). Let us first consider the infinitesimal world-sheet 
element of the $a$-th vortex loop, which has the form 

$$d\sigma_{\mu\nu}\left(x^a(\xi)\right)=\varepsilon^{\alpha\beta}
(\partial_\alpha x_\mu^a(\xi))
(\partial_\beta x_\nu^a(\xi))d^2\xi,$$
where $\alpha,\beta=1,2$.
Analogously to the 3D case, it is reasonable to introduce the 
center-of-mass coordinate (position) 
of the world-sheet $y_\mu^a\equiv\int d^2\xi
x_\mu^a(\xi)$. 
The full world-sheet coordinate can be respectively decomposed as follows 
$x_\mu^a(\xi)=y_\mu^a+z_\mu^a(\xi)$ with the vector $z_\mu^a(\xi)$ 
describing the shape of the $a$-th vortex loop world-sheet. 
Then, substituting 
into Eq.~(\ref{12}) instead of $\Sigma_{\mu\nu}$ the vorticity 
tensor current of the vortex loop gas~(\ref{13}), we can perform the 
summation over the grand canonical ensemble of the vortex loops as follows:

$$
1+\sum\limits_{N=1}^{\infty}\frac{\zeta'^N}{N!}\left(\prod\limits_{i=1}^{N}
\int d^4y^i\int {\cal D} z_\rho^i(\xi)\mu'\left[z^i\right]\right)
\sum\limits_{n_a=\pm 1}^{}\exp\left\{i\pi \sum\limits_{a=1}^{N}
n_a\int d\sigma_{\mu\nu}(z^a(\xi))h_{\mu\nu}(x^a(\xi))\right\}=$$

$$
=1+\sum\limits_{N=1}^{\infty}\frac{\left(2\zeta'\right)^N}{N!}\left\{
\int d^4y \int {\cal D}z_\rho(\xi)\mu'[z]\cos\left(\pi
\int d\sigma_{\mu\nu}(z(\xi))h_{\mu\nu}(x(\xi))\right)\right\}^N.$$
Here, $\mu'$ is a certain rotation- and translation invariant measure 
of integration over shapes of the world-sheets of the vortex loops .
Employing now the dilute gas approximation, we can expand $h_{\mu\nu}$ 
up to the first order in $a'/L'$ ({\it cf.} Eq.~(A.1)), 
where $a'$ stands for the typical value of
$\left|z^a(\xi)\right|$'s 
(vortex loops sizes), which are much smaller than the typical value 
$L'$ of 
$\left|y^a\right|$'s (distances between vortex loops)~\footnote{Similarly 
to the 3D case, the approximation $a'\ll L'$ means that the vortex loops 
are short living objects.}.
This yields 

$$
\int {\cal D}z_\rho(\xi)\mu'[z]
\cos\left(\pi
\int d\sigma_{\mu\nu}(z(\xi))h_{\mu\nu}(x(\xi))\right)\simeq 
\int {\cal D}z_\rho\mu'[z]\cos\left(
\frac{\pi}{L'}n_\lambda h_{\mu\nu}(y)P'_{\mu\nu, \lambda}[z]\right)=$$

$$
=\sum\limits_{n=0}^{\infty}\frac{(-1)^n}{(2n)!}\left(\frac{\pi}{L'}
\right)^{2n}n_{\lambda_1}h_{\mu_1\nu_1}(y)\cdots n_{\lambda_{2n}}
h_{\mu_{2n}\nu_{2n}}(y)\int {\cal D}z_\rho\mu'[z]
P'_{\mu_1\nu_1, \lambda_1}[z]\cdots 
P'_{\mu_{2n}\nu_{2n}, \lambda_{2n}}[z],\eqno (C.1)$$
where $n_\lambda\equiv y_\lambda/|y|$ and  
$P'_{\mu\nu, \lambda}[z]\equiv\int d\sigma_{\mu\nu}(z(\xi))
z_\lambda(\xi)$. Due to the rotation- and translation invariance 
of the measure $\mu'[z]$, the last average has the form 

$$
\int {\cal D}z_\rho\mu'[z]
P'_{\mu_1\nu_1, \lambda_1}[z]\cdots 
P'_{\mu_{2n}\nu_{2n}, \lambda_{2n}}[z]=$$

$$
=\frac{\left(a'^3\right)^{2n}}{(2n-1)!!}\left[\hat 1_{\mu_1\nu_1, 
\mu_2\nu_2}\delta_{\lambda_1\lambda_2}
\cdots \hat 1_{\mu_{2n-1}\nu_{2n-1}, \mu_{2n}\nu_{2n}}
\delta_{\lambda_{2n-1}\lambda_{2n}}+{\,}
{\rm permutations}\right].$$
Substituting this expression into Eq.~(C.1), we finally obtain 

$$
\int {\cal D}z_\rho(\xi)\mu'[z]
\cos\left(\pi
\int d\sigma_{\mu\nu}(z(\xi))h_{\mu\nu}(x(\xi))\right)\simeq
\sum\limits_{n=0}^{\infty}\frac{(-1)^n}{(2n)!}\left(
\frac{\pi a'^3}{L'}\right)^{2n}\left|h_{\mu\nu}(y)\right|^{2n}=
\cos\left(\frac{\left|h_{\mu\nu}(y)\right|}{\Lambda'^2}\right),$$
where we have introduced a new UV momentum cutoff 
$\Lambda'\equiv\sqrt{\frac{L'}{\pi a'^3}}$, which is much larger 
than $1/a'$. This yields the desired Eq.~(\ref{14}).

\end{document}